\pdfoutput=1

\documentclass[11pt,a4paper]{article}

\usepackage[utf8]{inputenc}
\usepackage[T1]{fontenc}
\usepackage{lmodern}
\usepackage[margin=1in]{geometry}
\usepackage{graphicx}
\usepackage{booktabs}
\usepackage{hyperref}
\usepackage{amsmath}
\usepackage{amssymb}
\usepackage{amsthm}
\usepackage[round]{natbib}
\usepackage{xcolor}
\usepackage{microtype}
\usepackage{algorithm}
\usepackage{algpseudocode}
\usepackage{appendix}
\usepackage{tabularx}

\newtheorem{definition}{Definition}

\hypersetup{
    colorlinks=true,
    linkcolor=blue,
    citecolor=blue,
    urlcolor=blue,
    pdftitle={Phantom Citations or Phantom Detectors? A Source-Level Re-Audit of Reference Integrity in AI Survey Papers},
    pdfauthor={H. Kemal Ilter}
}

\title{Phantom Citations or Phantom Detectors?\\A Source-Level Re-Audit of Reference Integrity in AI Survey Papers}

\author{
    H. Kemal \.{I}lter\thanks{ORCID: 0000-0002-6359-9976}\\
    \textit{Department of Management Information Systems}\\
    \textit{Bak{\i}r\c{c}ay University, \.{I}zmir, Turkey}\\
    \texttt{kemal.ilter@bakircay.edu.tr}
}

\date{November 2026 \\ \small Version 2. This version replaces and corrects version 1 (January 2026); see Section~\ref{sec:correction}.}

\begin{document}

\maketitle

\begin{abstract}
\textbf{Background:} Version~1 of this preprint reported that 17.0\% of citations in 50 recent survey papers on large language models (LLMs) could not be resolved to any digital object, and interpreted this ``Phantom Rate'' as evidence of endemic metadata hallucination. A re-examination showed that this figure was produced largely by the audit pipeline rather than by the surveyed literature.

\textbf{Methods:} We re-audited the same 50 arXiv survey versions, reading each bibliography from the authors' \LaTeX{} source instead of from text extracted from PDF files. Each of the 9{,}281 references was checked against the DOI system, arXiv, Crossref and OpenAlex under a strict identifier-extraction rule and a three-state outcome scheme that never counts a failed lookup as evidence of non-existence. Unresolved and partially matched references were searched again with an author comparison and adjudicated by hand.

\textbf{Results:} Four errors in version~1 account for its headline result and its interpretation: a greedy DOI pattern that captured text beyond valid identifiers, an extraction step that recovered no citations from 16 of the 50 papers, per-paper statistics that cannot be reproduced from the archived data, and a study period reported as 16 months that in fact spans four and a half. Of the 939 citations classified as phantoms in version~1, 731 correspond to references in the source bibliographies and 652 of these resolve to the cited work; none corresponds to a fabricated reference. The source-level audit identifies 13 fabricated references among 9{,}012 checkable scholarly references (0.14\%; 95\% CI 0.08--0.25\%). Nine are chimeric: they attach a real work's subject, and in two cases its correct arXiv identifier, to invented authors and an altered title or venue. The other four could not be traced to any publication. Eleven of the thirteen occur in two papers, and nine occur in papers from which version~1 extracted no citations.

\textbf{Conclusion:} Audits that read citations from PDF text can overstate reference fabrication by more than two orders of magnitude. Fabrication in this corpus is rare, but where it occurs it can pass both title-based and identifier-based checks. Automated author comparison flagged only two of the nine chimeric references, and detection depended on reading the entries.

\vspace{0.5em}
\end{abstract}

\section{Introduction}

Citations are the means by which a scholarly claim can be traced to its evidence \citep{merton1973sociology}. Survey papers occupy a particular position in this system: they condense large literatures into reference lists that later authors reuse, so an error in a widely read survey can travel further than an error in a single empirical study. The growing use of large language models (LLMs) in literature review has made the integrity of these reference lists a practical question, because generative models are known to produce references that do not exist \citep{alkaissi2023artificial,athaluri2023exploring,ji2023survey}.

Version~1 of this preprint addressed that question with an automated audit of 50 recent survey papers on LLMs and reported that 17.0\% of their citations could not be resolved to any digital object. It interpreted this figure as a stable ``equilibrium of decay'', attributed most failures to models that retrieve correct titles but invent identifiers, and projected the consequences with a model of irreversible error accumulation. A reader's question about how the 17.0\% related to the paper's own breakdown of failure modes prompted a re-examination of the pipeline, the data and the extracted citation strings. That re-examination showed that the result does not hold.

This version reports the re-audit and corrects the record. It makes three contributions. First, it documents the specific errors that produced version~1's figures and traces every citation that version~1 classified as a phantom to the corresponding reference in the authors' source files (Section~\ref{sec:correction}). Second, it provides a source-level estimate of reference fabrication in the same corpus, using a verification design in which every reference ends in an explicit outcome and in which an unanswered query is never treated as evidence that a work does not exist (Sections~\ref{sec:methods} and~\ref{sec:results}). Third, it describes the form that the remaining fabrications take. They are not inventions from nothing but recombinations of real bibliographic elements, a pattern with direct consequences for the design of verification tools (Section~\ref{sec:discussion}).

The correction carries a methodological lesson that extends beyond this paper. Automated citation audits are increasingly used to estimate the prevalence of AI-generated references. When such audits read references from PDF text, the measurement instrument introduces errors of the same appearance as the phenomenon it is meant to detect: identifiers broken by lost whitespace look like invented identifiers, and fragments of tables look like references to works that cannot be found. Without a design that separates these sources of error, an audit can report a large effect that belongs to the instrument rather than to the literature.

\section{Related Work}

\subsection{LLM Hallucination in Academic Contexts}

The tendency of LLMs to generate plausible but incorrect content has been documented extensively \citep{ji2023survey}. In academic writing it appears, among other forms, as references to publications that do not exist. Early reports based on ChatGPT \citep{alkaissi2023artificial,athaluri2023exploring} and subsequent commentary \citep{azamfirei2023large} drew attention to this behaviour; most such studies examine references that a model generates on request rather than references that appear in published work.

\subsection{Link Decay and Reference Rot}

A separate literature examines whether references that were correct when published remain usable. \citet{hennessey2013cross} studied the decay of web links across disciplines and the effectiveness of mitigation techniques, and \citet{klein2014scholarly} found that a substantial share of web-based references in scholarly articles suffer from reference rot. These studies concern the loss of access to genuine sources over time; the present paper concerns whether the cited sources exist at all.

\subsection{Persistent Identifiers and Open Metadata}

The DOI system provides persistent identifiers for scholarly objects \citep{paskin2010digital}, and Crossref maintains openly accessible metadata for the works registered through it \citep{hendricks2020crossref}. OpenAlex offers an open index of scholarly works, authors and venues that also covers material without a Crossref DOI \citep{priem2022openalex}. Verification pipelines, including the one described here, depend on these infrastructures, and their coverage limits what any automated audit can establish (Section~\ref{sec:limitations}).

\section{Methods}
\label{sec:methods}

\subsection{Corpus}

The corpus consists of the same 50 arXiv survey papers analysed in version~1, identified by the query \texttt{ti:"Survey" AND (ti:"Large Language Models" OR ti:"Generative AI")} and submitted between 8 September 2025 and 20 January 2026. Each paper was retrieved at the exact version audited in version~1. For every paper, the \LaTeX{} source distributed by arXiv was downloaded and the bibliography was read from it, in the following order of preference: a compiled bibliography file (\texttt{.bbl}; 38 papers), a \texttt{thebibliography} environment in the main source (2 papers), the cited entries of BibTeX files (\texttt{.bib}; 9 papers), and a plain enumerated reference list (1 paper). This procedure yielded 9{,}281 references from all 50 papers.

\subsection{Identifier Extraction}

Digital Object Identifiers were extracted with a pattern that does not allow whitespace, braces, quotation marks, backslashes or commas in the DOI suffix. Text attached to the end of an identifier beginning with \texttt{URL} or \texttt{http} was removed, as were trailing punctuation marks and unbalanced closing parentheses. Strings whose suffix was shorter than three characters, such as the truncated form \texttt{10.48550/ar}, were rejected as incomplete rather than treated as identifiers. arXiv identifiers were taken from explicit \texttt{arXiv:} prefixes, arXiv URLs, CoRR \texttt{abs/} references and arXiv DOIs.

\subsection{Verification Sequence}

Each reference was checked in the following sequence.
(i) A cited DOI was resolved through the DOI Handle System, which covers all registration agencies. If it resolved, the title registered for it was retrieved by content negotiation and compared with the cited title, so that an identifier belonging to a different work could be detected.
(ii) A cited arXiv identifier was checked through the arXiv API. Because the API reports only the latest title of a paper, the title of version~1 was also compared, which accounts for papers renamed after the cited version.
(iii) If neither identifier confirmed the reference, the citation was searched in Crossref (bibliographic query on the full citation) and in OpenAlex (title search) \citep{hendricks2020crossref,priem2022openalex}.
All responses were cached so that the audit can be re-run on identical evidence.

\subsection{Matching Rule}

Titles were compared with the Levenshtein similarity ratio and the thresholds of version~1: a reference matched a record if the similarity was at least 85\%, and partially matched it between 50\% and 85\%. Three safeguards were added after inspection of a pilot sample. A match was accepted only if every substantive word of the cited title (five or more letters, excluding common function words) had a close counterpart in the record's title, which prevents near-identical strings with a different key term from being accepted. Added or dropped subtitles were handled by comparing each title with the segments of the other title delimited by a colon or question mark, provided the segment contained at least three substantive words. When a title could be extracted from the reference, the record was compared with the title alone and never with the full citation string, so that venue names could not produce matches with proceedings volumes.

\subsection{Outcome Categories}

Every reference received exactly one outcome (Table~\ref{tab:outcomes}). The scheme distinguishes references that were checked and not found from references that could not be checked, and it reports web pages, software, datasets and model cards separately from scholarly works.

\begin{table}[htbp]
\centering
\caption{Outcome categories of the source-level audit.}
\label{tab:outcomes}
\small
\begin{tabularx}{\textwidth}{lX}
\toprule
Outcome & Definition \\
\midrule
Verified & The cited identifier or a title search confirms the cited work. \\
Metadata error & The work exists, but the cited DOI or arXiv identifier does not resolve or belongs to another work. \\
Partial match & The best match has a similarity between 50\% and 85\%. \\
Identifier mismatch & The cited identifier resolves to a different work and the cited title is not found. \\
Not found & All lookups were completed and no record reached 50\% similarity. \\
Grey literature & Web page, software, dataset, model card or similar; not assessed as a scholarly work. \\
Not checked & At least one lookup failed, or the entry contained too little text to search. \\
\bottomrule
\end{tabularx}
\end{table}

\subsection{Second Pass with Author Comparison}

All references classified as partial match, not found or identifier mismatch were searched again by title in arXiv and in OpenAlex, which returns author lists. DBLP was also queried but refused automated requests during the study period and was not used. For the best-matching record, the surnames of its first three authors were compared with the cited author list. The combination of title similarity and author agreement assigned each reference to one of five tiers: chimera suspect (similar or identical title, different authors), not found, not checked, title variant (same authors, different title) and real (title match with agreeing authors). Author comparison was introduced because the fabrications identified in the pilot combined real titles or near-real titles with authors who did not write them.

\subsection{Manual Adjudication}

Every reference in the five tiers (326 in total) was assigned one of eight verdicts. References whose status was not evident from the record found by the second pass, 46 in all, were checked against publisher pages, the ACL Anthology, arXiv, OpenReview and general web search; the remaining 280 were judged from the record found by the second pass, the cited entry and bibliographic knowledge. Verdicts were prepared with the assistance of an AI research assistant (Claude, Anthropic) and reviewed by the author. A reference was classified as \emph{fabricated} if no publication matching its cited title and authors could be found and, where its elements could be traced, they belonged to different real publications. A reference was classified as \emph{real} if the cited work was located, including works absent from the indexes queried automatically. The remaining categories were recorded as grey literature, metadata error (a real work cited with a wrong identifier) and corrupted entry (a reference to a real work whose source entry was damaged, for example by unexpanded macros). Two categories were added during adjudication. A reference was classified as \emph{likely fabricated} if no trace of the cited work could be found under its title, its authors or its venue, and no real work could be identified of which it was a corrupted version. A reference was classified as \emph{altered metadata} if it pointed to a real, identifiable work but changed its title or an author's name; such references are counted with the metadata errors, not with the fabrications. The fabrication rate is reported for chimeric and likely fabricated references together and, as a lower bound, for chimeric references alone. A reference whose existence could neither be confirmed nor excluded was recorded as \emph{unresolved} and is reported as a sensitivity bound. Entries reduced by extraction to an author list, a venue or an affiliation were recorded as \emph{parser errors}. The per-reference verdicts, notes and the basis of each verdict are archived with the audit code.

\subsection{Crosswalk to Version 1}

To trace what version~1 measured, each of its 5{,}514 extracted citation strings was matched to a reference in the same paper's source bibliography. Because version~1's strings frequently lost their spaces during PDF extraction, both strings were reduced to lower-case alphanumeric characters, and the extracted title of the source reference (or the first 60 characters of the reference when no title was available) was located in the version~1 string by partial fuzzy matching. A string was assigned to a reference when the partial similarity reached 85\%.

\subsection{Statistical Analysis and Reproducibility}

Proportions are reported with 95\% Wilson score intervals (Appendix~\ref{app:ci}). The audit code, the cached evidence, the per-reference outcomes, the adjudication records and the crosswalk are archived in Zenodo at \url{https://doi.org/10.5281/zenodo.24587492}, together with the version~1 classifications on which the crosswalk is based. The arXiv \LaTeX{} sources of the surveyed papers are not redistributed; the code retrieves them from arXiv. The materials of version~1 remain available at \url{https://doi.org/10.17605/OSF.IO/T8S53}; the Zenodo record supersedes them.

\section{Results}
\label{sec:results}

\subsection{Coverage}

Reading bibliographies from source files recovered 9{,}281 references from the 50 papers, compared with 5{,}514 citation strings recovered from 34 papers in version~1. Every paper yielded a bibliography.

\subsection{Automated Outcomes}

Table~\ref{tab:auto} reports the outcomes of the automated audit before the second pass and manual adjudication. Of the 9{,}281 references, 8{,}645 (93.1\%) were verified automatically.

\begin{table}[htbp]
\centering
\caption{Automated outcomes of the source-level audit ($N = 9{,}281$ references, 50 papers).}
\label{tab:auto}
\begin{tabular}{lrr}
\toprule
Outcome & Count & Share \\
\midrule
Verified & 8{,}645 & 93.1\% \\
Partial match & 277 & 3.0\% \\
Grey literature & 261 & 2.8\% \\
Metadata error & 41 & 0.4\% \\
Not found & 37 & 0.4\% \\
Identifier mismatch & 12 & 0.1\% \\
Not checked & 8 & 0.1\% \\
\bottomrule
\end{tabular}
\end{table}

\subsection{Adjudicated Fabrications}

\begin{figure}[htbp]
\centering
\includegraphics[width=\textwidth]{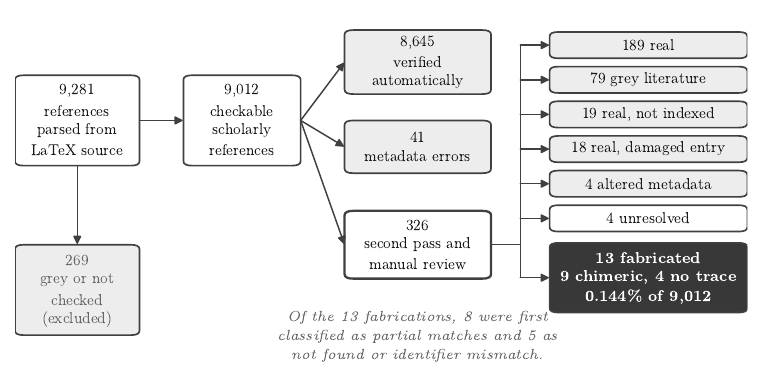}
\caption{From parsed references to adjudicated fabrications. Of 9{,}281 references parsed from the \LaTeX{} sources of 50 papers, 269 were grey literature or could not be checked; 8{,}645 of the remaining 9{,}012 were verified automatically and 41 were real works with a faulty identifier. The 326 references sent to the second pass and manual review contained 13 fabrications, 0.144\% of the checkable references.}
\label{fig:flow}
\end{figure}

Figure~\ref{fig:flow} summarises the path from parsed references to adjudicated fabrications. Table~\ref{tab:adjudication} reports the verdicts for the 326 references in the review queue. Most were real works: 189 were located as cited or under a later title, 19 were real but absent from the indexes queried automatically, and 18 were real works whose entries had been damaged in extraction. A further 79 were grey literature that the automated rules had not recognised, most often blog posts, model cards, technical reports and theses.

\begin{table}[htbp]
\centering
\caption{Adjudicated verdicts for the 326 references in the review queue, by second-pass tier. Chim.\ = chimeric; Fabr.\ = likely fabricated; Alt.\ = altered metadata; Not idx.\ = real but not indexed; Parser = real work with a damaged entry; Unres.\ = unresolved.}
\label{tab:adjudication}
\small
\begin{tabular}{lrrrrrrrrr}
\toprule
Tier & Chim. & Fabr. & Alt. & Real & Not idx. & Grey & Parser & Unres. & Total \\
\midrule
1 Chimera suspect & 2 & 2 & 2 & 39 & 5 & 14 & 2 & 1 & 67 \\
2 Not found & 5 & 2 & 2 & 43 & 13 & 57 & 13 & 3 & 138 \\
3 Not checked & 0 & 0 & 0 & 12 & 1 & 7 & 3 & 0 & 23 \\
4 Title variant & 2 & 0 & 0 & 54 & 0 & 1 & 0 & 0 & 57 \\
5 Real & 0 & 0 & 0 & 41 & 0 & 0 & 0 & 0 & 41 \\
\midrule
Total & 9 & 4 & 4 & 189 & 19 & 79 & 18 & 4 & 326 \\
\bottomrule
\end{tabular}
\end{table}

Adjudication identified nine chimeric and four likely fabricated references among 9{,}012 checkable scholarly references: 13 in total, a rate of 0.144\% (95\% CI 0.084--0.247\%). Counting chimeric references alone gives 0.100\% (95\% CI 0.053--0.190\%). Excluding from the denominator the 79 references that adjudication reclassified as grey literature changes the rate only in the third decimal (0.146\%). Four references remained unresolved; if all four were fabricated, the rate would be 0.189\% (95\% CI 0.118--0.302\%), still two orders of magnitude below the rate reported in version~1.

The automated outcome categories located the fabrications poorly. Only five of the 13 had been classified as not found or identifier mismatch, the categories the first pass reported as candidate fabrications; the other eight had been classified as partial matches, because their titles resembled those of real works. The candidate rate of the first pass (49 references, 0.54\%) therefore both overstates fabrication, since most candidates were real, and misses most of the fabrications it is meant to capture. The verified cases share a pattern, which we term a \emph{chimeric reference}.

\begin{definition}[Chimeric reference]
A reference whose bibliographic elements (subject, title, authors, venue, identifier) are individually plausible and partly drawn from real publications, but which in combination describe a publication that does not exist.
\end{definition}

Three cases illustrate the pattern. The first cites ``REPE: Representation and Prediction-Level Alignment for Multilingual Intent Detection and Slot Filling'' by authors named Zhang, Chen and Zhao, in the EMNLP 2024 main proceedings, with an ACL Anthology DOI. REPE exists, but it was published by Zhu, Cheng, Chen, Zhuang, Huang and Zou as ``Code-Switching Can be Better Aligners: Advancing Cross-Lingual SLU through Representation-Level and Prediction-Level Alignment'' in the ACL 2024 short papers, under a different DOI. The second cites ``ContrastiveMix: Overcoming Code-mixing Dilemmas in Multilingual Spoken Language Understanding'' by Lin, Chen and Liu as a NAACL 2024 long paper; the real ContrastiveMix is a NAACL 2024 short paper by Do, Lee and Hwang on cross-lingual information retrieval. The third cites ``Large Language Models Have General Value-Based Biases'' by Schramowski, Ceyda Turan and Nora Andersen in \emph{Patterns} (2022) with a \emph{Patterns} DOI; the work it resembles is ``Large pre-trained language models contain human-like biases of what is right and wrong to do'' by Schramowski, Cigdem Turan, Nico Andersen, Rothkopf and Kersting in \emph{Nature Machine Intelligence} (2022). In each case the identifier has the correct form and the correct registrant prefix for the cited venue, which is why the reference survives casual inspection.

The remaining chimeric references vary this pattern. Two carry the correct arXiv identifier of the work they distort. CodeMixBench is cited with the identifier of the preprint by Sheokand and Sawant (arXiv:2505.05063) but with two added co-authors, an altered title and the ICON conference as venue; Self-Refine is cited with the identifier of the NeurIPS 2023 paper by Madaan and colleagues (arXiv:2303.17651) but under an invented author list and a title extended by ``with GPT-4''. For these references, identifier resolution confirms the citation rather than exposing it. Two further cases move a real work to another venue and change its authors: the MultiCoNER dataset paper of Malmasi and colleagues (COLING 2022) is cited with added co-authors, an extended title and NAACL as venue, and the Marathi corpus paper of Chavan and colleagues (Findings of IJCNLP-AACL 2023) is cited with different authors, a different title and EACL as venue. In 2511.07448v1, the Scideator and Virtual Scientists systems are attributed to generic name pairs (``Smith, George; Huang, Lisa'' and ``Johnson, Peter; Davis, Emily'') and to \emph{Transactions on Machine Learning Research}, in which neither was published.

Invented authors also recur across entries. The two co-authors added to the CodeMixBench entry appear again on a code-mixed machine translation shared-task paper attributed to Dhar and colleagues, for which no record exists. This entry and an IndicTrans demonstration paper attributed to an author named ``The AI Guy'' are the two likely fabricated references in 2510.07037v5; the other two, ``CodeScientist'' and ``Explanations as Programs'', both attributed to \emph{Transactions on Machine Learning Research}, are in 2511.07448v1.

\subsection{Concentration}

Fabricated references are concentrated in a few papers (Table~\ref{tab:concentration}). Seven occur in 2510.07037v5 and four in 2511.07448v1; 2512.06874v1 and 2509.10697v1 contain one each, and the other 46 papers contain none. The two papers with most fabrications also have many other problems: 2511.07448v1 has the largest number of metadata errors in the corpus (six of 41), and 2510.07037v5 has the largest number of partial matches (24). Problems with references therefore cluster within papers rather than being spread thinly across the literature. Three of the four papers, holding nine of the thirteen fabrications, are among the 16 from which version~1 extracted no citations.

\begin{table}[htbp]
\centering
\caption{Papers containing references adjudicated as fabricated. Papers are identified by arXiv identifier and version. Refs = references parsed from the \LaTeX{} source. Read in v1 = version~1 extracted citations from the paper. }
\label{tab:concentration}
\small
\begin{tabular}{lrrrrc}
\toprule
Paper & Refs & Chimeric & Likely fabricated & Total & Read in v1 \\
\midrule
2510.07037v5 & 336 & 5 & 2 & 7 & No \\
2511.07448v1 & 194 & 2 & 2 & 4 & Yes \\
2512.06874v1 & 25 & 1 & 0 & 1 & No \\
2509.10697v1 & 315 & 1 & 0 & 1 & No \\
\midrule
Total & 870 & 9 & 4 & 13 & \\
\bottomrule
\end{tabular}
\end{table}

\subsection{Metadata Errors}

The 41 metadata errors are real works cited with a faulty identifier. In 30 the identifier resolves to a different work (19 DOIs and 11 arXiv identifiers), and in 5 a cited DOI does not resolve. Four concern papers renamed after their first arXiv version, cited with an identifier that points elsewhere. Two carry ACM Digital Library identifiers with the prefix 10.5555, which ACM displays as DOIs but which are not registered and therefore never resolve; these are errors of the infrastructure rather than of the citing authors. Like fabrications, metadata errors cluster: four papers account for 21 of the 41. Adjudication added four references with altered metadata, which point to real works under a changed title or author name. The direct preference optimisation paper of Rafailov and colleagues is cited with an altered title and the first author's given name changed, and a refusal-training paper by Cao with ``More Trustworthy'' inserted into its title. One entry carries the correct arXiv identifier (2407.08836) of a paper on fault diagnosis in power grids by Jing and Rahman but gives it a different title and different initials. The fourth attributes the 2023 \emph{Nature} feature ``The promise and peril of generative AI'', written by Stokel-Walker and Van Noorden, to ``Jo, A.''; the name matches the credit of the feature's illustrator, Pawe{\l} Jo\'{n}ca, which suggests an error inherited from automatically harvested metadata rather than an invention. None of the four is counted as a fabrication.

\section{Correction to Version 1}
\label{sec:correction}

This section documents the errors in version~1 and their consequences. All statements refer to the version~1 data and code archived at \url{https://doi.org/10.17605/OSF.IO/T8S53}.

\subsection{Identifier Extraction}

Version~1 extracted DOIs with the pattern \texttt{10\textbackslash.\textbackslash d\{4,9\}/[-.\_;()/:A-Za-z0-9]+}, which admits letters, dots, colons and slashes after a valid DOI. When PDF extraction removed the space between a DOI and the following text, the pattern captured both; for example, \texttt{10.1145/3544548.3581340.URLhttps://doi.org/...} was submitted for resolution as a single identifier, failed, and was labelled a hallucinated identifier. Of the 154 citations classified as ``Broken Links'', 116 contain such appended text, 33 are truncated identifiers such as \texttt{10.48550/ar}, and 5 have the form of complete arXiv DOIs. The note attached to these records (``confirmed hallucination'') was produced by a version of the pipeline that is not included in the archived code, so this classification cannot be reproduced from the published materials.

\subsection{Incomplete Coverage of the Corpus}

Version~1 reported results for 50 papers, but its extraction step recovered no citations from 16 of them. The 5{,}514 citations therefore come from 34 papers. The 16 papers without extracted citations contain 3{,}079 references in their source files, a third of the corpus, and they include three papers that contain nine of the thirteen references adjudicated as fabricated or likely fabricated.

\subsection{Study Period}

Version~1 described the corpus as published between September 2024 and January 2026 and interpreted a flat regression slope ``over 16 months'' as evidence of a stable equilibrium. The submission dates of the 50 papers range from 8 September 2025 to 20 January 2026, a period of about four and a half months. The temporal analysis therefore covered a much shorter window than stated, in addition to the problems described below.

\subsection{Per-Paper Statistics}

The paper-level statistics in version~1 (its Table~4: mean 16.5\%, median 12.8\%, maximum 58.8\%; its Figure~3; and the temporal regression in its Section~4.4) cannot be reproduced from the archived data. The reported maximum of 58.8\% corresponds to per-paper rates stored before the reclassification step, whereas the corpus-level rate of 17.0\% uses the counts after it, and the 16 papers without citations enter the archived per-paper rates as zero.

\subsection{Interpretation of Unresolved Citations}

Section~5.3 of version~1 argued that 17.0\% was a lower bound because some ``Unknown'' citations were probably phantoms. Given the misclassifications above, the opposite holds: 17.0\% is best read as the rate of automatic phantom classifications produced by the version~1 pipeline, an upper bound on flags rather than a lower bound on references that lead nowhere.

\subsection{Crosswalk}

\begin{figure}[htbp]
\centering
\includegraphics[width=\textwidth]{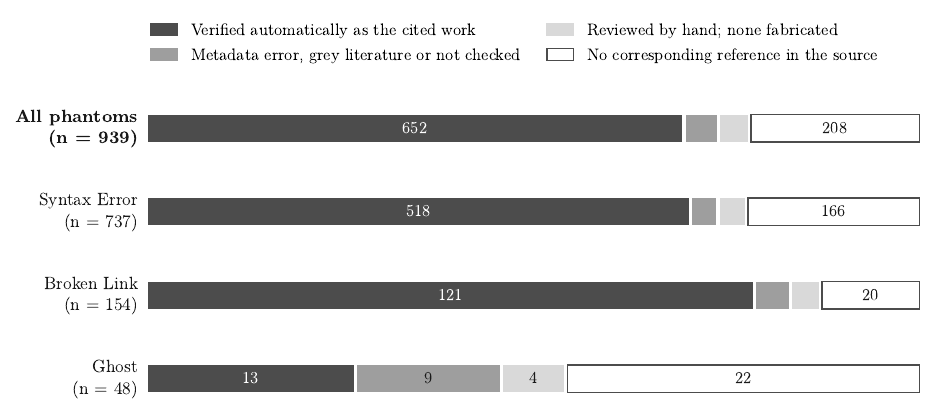}
\caption{What the version~1 phantoms were. Each bar traces the citations version~1 classified as phantoms, by subcategory, to the source-level audit. Of all 939, 652 were verified automatically as the cited work, 42 were real works with a metadata error, grey literature or not checked, 37 were reviewed by hand and none was fabricated, and 208 had no corresponding reference in the source bibliography.}
\label{fig:v1}
\end{figure}

Table~\ref{tab:crosswalk} traces every version~1 citation to the source-level audit, and Figure~\ref{fig:v1} shows the outcome for the citations version~1 classified as phantoms. Of the 939 phantoms, 731 correspond to a reference in the source bibliography; 652 of these were verified automatically. The 208 phantoms without a corresponding reference are consistent with fragments of tables, captions and broken lines that version~1 treated as citations. None of the version~1 phantoms corresponds to a reference adjudicated as fabricated: version~1 flagged real works and did not detect the fabrications, nine of which occur in papers from which it extracted no citations.

\begin{table}[htbp]
\centering
\caption{Version~1 categories traced to source-level outcomes. Ver.\ = verified; Other = metadata error, partial match, grey literature or not checked; Cand.\ = not found or identifier mismatch before adjudication; Fabr.\ = adjudicated as fabricated; None = no corresponding reference in the source bibliography.}
\label{tab:crosswalk}
\small
\begin{tabular}{lrrrrrr}
\toprule
Version 1 category & $n$ & Ver. & Other & Cand. & Fabr. & None \\
\midrule
Valid & 2{,}259 & 1{,}990 & 112 & 4 & 0 & 153 \\
Sloppy (recovered) & 536 & 453 & 34 & 0 & 0 & 49 \\
Unknown & 1{,}780 & 1{,}172 & 89 & 9 & 0 & 510 \\
Phantom: Syntax Error & 737 & 518 & 45 & 8 & 0 & 166 \\
Phantom: Broken Link & 154 & 121 & 10 & 3 & 0 & 20 \\
Phantom: Ghost & 48 & 13 & 13 & 0 & 0 & 22 \\
\bottomrule
\end{tabular}
\end{table}

\subsection{Claims Withdrawn}

The following claims of version~1 are withdrawn: the 17.0\% Phantom Rate as a measure of citations that cannot be resolved; the shares of pure hallucinations (5.1\%), hallucinated identifiers (16.4\%) and parsing failures (78.5\%); the description of the study period as September 2024 to January 2026 and the interpretation of a flat temporal trend as an ``equilibrium of decay''; the half-life and equilibrium projections derived from a phantom rate of 0.17 (version~1, Section~5.2 and Appendix~E); and the argument that 17.0\% is a lower bound. The recommendation that submission systems verify references automatically (version~1, Section~5.4) is retained, subject to the qualifications in Section~\ref{sec:discussion}.

\section{Discussion}
\label{sec:discussion}

\subsection{The Instrument as a Source of Apparent Hallucination}

The central finding of the re-audit concerns measurement. Every component of version~1's estimate that looked like evidence of fabrication had an origin in the pipeline: identifiers that failed because extraction had joined them to the following text, titles that could not be matched because whitespace had been lost, and ``references'' that were fragments of tables. These errors are not idiosyncratic. Any audit that reads references from PDF text is exposed to them, and they produce exactly the signatures, unresolvable identifiers and unmatched titles, that such audits interpret as hallucination. Reading bibliographies from source files, where they are available, removes most of this error; where they are not, an audit should report extraction failures separately and verify a sample of its positive findings by hand before reporting a prevalence.

\subsection{Chimeric References}

The fabrications that remain are informative because of their form. Version~1 described LLMs as ``lazy research assistants'' that retrieve correct titles and invent identifiers. The verified cases suggest a different mechanism: recombination. A real contribution (REPE, ContrastiveMix, the moral-direction work of Schramowski and colleagues) is reassembled with authors, a title, a venue and an identifier that are each plausible for that topic and venue but do not belong together. The result passes the checks a reader or a simple tool applies: the acronym is real, the venue is real, and the DOI has the correct registrant prefix. The references themselves cannot show how they were produced. Recombination of this kind is consistent with generation by a language model, but an author assembling entries from memory, or a reference manager filled from a faulty source, could produce some of the same errors. The recurrence of the same invented co-authors on different entries, and the generic name pairs in 2511.07448v1, make independent human slips a less plausible explanation for the clusters in these two papers; they do not establish the cause.

\subsection{Implications for Verification Tools}

Two design consequences follow. First, existence checks must be distinguished from support checks, and both must distinguish ``not found'' from ``not checked''; a tool that reports every unanswered query as a possible fabrication will accuse real but unindexed work, as some current systems do. Second, neither title similarity nor author comparison is sufficient on its own. A chimeric reference can score between 50\% and 85\% against its real counterpart and be classified as a minor discrepancy. Comparing authors helps only when the title search finds the real work: in this audit the tier combining title similarity with author disagreement held 67 references, of which four were fabricated and only two chimeric, while five chimeric references were not found at all, because their altered titles no longer led to the real work, and two were classed as title variants. Identifier resolution can mislead in the opposite direction, since two chimeric references carried the correct arXiv identifier of the work they distort. Submission-time verification remains desirable, but it should report its evidence and its uncertainty rather than a single pass or fail.

\subsection{Propagation}

Version~1 modelled the accumulation of citation errors across generations of surveys as a ratchet. With an observed fabrication rate of the order of 0.1\%, the projections of version~1 no longer apply. Whether chimeric references, once published, are copied into later work remains an open empirical question; their distinctive combinations of title, authors and identifier make them suitable markers for tracing such copying.

\subsection{Limitations}
\label{sec:limitations}

The audit covers arXiv preprints with available \LaTeX{} sources in one subfield and one period. Automated verification depends on the coverage of the DOI system, Crossref, OpenAlex and arXiv; references to workshop papers, theses and technical reports are often absent from these indexes, which is why adjudication by hand was required. The audit establishes whether a cited work exists, not whether it supports the statement for which it is cited. No verdict was checked independently by a second coder, and 280 of the 326 verdicts rest on the record found automatically and on bibliographic knowledge rather than on an individual search; these are concentrated in the title-variant and real tiers, where author lists agreed, and among grey literature. Four references remained unresolved. The second pass also depended on services that throttled or refused automated queries: DBLP could not be used, and arXiv title search stopped partway through, so later references were searched in OpenAlex alone. Finally, the rate counts only references that could be shown to be fabricated; a chimeric reference whose elements all happen to belong to unindexed work would be adjudicated as not found rather than fabricated.

\section{Conclusion}

The 17\% figure reported in version~1 was an artifact of how citations were extracted and classified, not a property of the surveyed literature. Read from their source files, the reference lists of 50 recent LLM surveys contain fabricated references at a rate of about 0.14\%, concentrated in a few papers. These fabrications are rare but not trivial: they recombine real bibliographic elements into publications that do not exist, and they are designed, in effect, to pass the checks that readers and simple tools apply. Measuring them requires an instrument whose own errors are smaller than the effect it seeks; the correction of version~1 shows how easily that condition can fail.

\section*{Data Availability}

The re-audit code, cached API responses, per-reference outcomes, adjudication records and the crosswalk to version~1 are available in Zenodo at \url{https://doi.org/10.5281/zenodo.24587492} (code under the MIT licence, data under CC~BY~4.0). The version~1 materials remain at \url{https://doi.org/10.17605/OSF.IO/T8S53} and are superseded by the Zenodo record.

\section*{Acknowledgments}

I thank Rafa{\l} Wrona for a question about version~1 that prompted this re-examination, and Mohamed Hani (University of Granada) for discussions on reference verification in AI-assisted writing tools. This work used the DOI Handle System, Crossref, OpenAlex and the arXiv API; I thank the organisations that maintain this open infrastructure.

\bibliographystyle{apalike}
\bibliography{references}

\appendix

\section{Verification Algorithm}
\label{app:algorithm}

\begin{algorithm}[htbp]
\caption{Source-level reference verification (one reference $r$)}
\begin{algorithmic}[1]
\Require reference $r$ with cited title $t$, raw text, optional DOI $d$ and arXiv identifier $a$
\If{$d \neq \emptyset$}
  \State $e \gets$ \Call{HandleExists}{$d$}
  \If{$e = $ yes}
    \State $t_d \gets$ \Call{RegisteredTitle}{$d$}; $s \gets$ \Call{Score}{$t, t_d$}
    \If{$s \geq 85$} \Return Verified \EndIf
    \If{$s < 50$} mark identifier mismatch \EndIf
  \ElsIf{$e = $ no} mark identifier failure \EndIf
\EndIf
\If{$a \neq \emptyset$}
  \State $s \gets \max(\Call{Score}{t, \text{latest title}}, \Call{Score}{t, \text{v1 title}})$
  \If{$s \geq 85$} \Return Verified, or Metadata error if a DOI failure or mismatch was marked \EndIf
\EndIf
\State $s \gets$ best \Call{Score}{} over Crossref and OpenAlex results
\If{$s \geq 85$} \Return Verified, or Metadata error if an identifier failure or mismatch was marked
\ElsIf{identifier mismatch marked} \Return Identifier mismatch
\ElsIf{$r$ is a web page, software or dataset} \Return Grey literature
\ElsIf{$s \geq 50$} \Return Partial match
\ElsIf{any lookup failed} \Return Not checked
\Else{} \Return Not found
\EndIf
\end{algorithmic}
\end{algorithm}

\textsc{Score} is the Levenshtein similarity ratio with the safeguards described in the Matching Rule subsection of Section~\ref{sec:methods}.

\section{Confidence Intervals}
\label{app:ci}

Proportions are reported with the Wilson score interval,
\begin{equation}
\text{CI}_{95\%} = \frac{\hat{p} + \frac{z^2}{2n} \pm z\sqrt{\frac{\hat{p}(1-\hat{p})}{n} + \frac{z^2}{4n^2}}}{1 + \frac{z^2}{n}},
\end{equation}
with $z = 1.96$.

\end{document}